# Designing Virtual Soundscapes for Alzheimer's Disease Care

Frédéric Voisin[1]

[1] Freelance
fred@fredvoisin.com

**Abstract.** Sound environment is a prime source of conscious and unconscious information which allows listeners to place themselves, to communicate, to feel, to remember. The author describes the process of designing a new audio interactive apparatus for Alzheimer's care, in the context of an active multidisciplinary research project led by the author in collaboration with a longterm care centre (EHPAD) in Burgundy (France), a geriatrician, a gerontologist, psychologists and caregivers. The apparatus, named *Madeleines Sonores* in reference to Proust's *madeleine*, have provided virtual soundscapes sounding for a year for 14 elderly people hosted in the dedicated Alzheimer's unit of the care centre, 24/7. Empiric aspects of sonic interactivity are discussed in relation to dementia and to the activity of caring. Scientific studies are initiated to evaluate the benefits of such a disposal in Alzheimer's disease therapy and in caring dementia.

## 1 Introduction

Nowadays, Alzheimer's disease (AD) is the main cause of dementia. AD usually starts slowly and gradually worsens over time with brain damages and main effects on memory, cognition and behiavior : short-term and procedural memories are progressively affected until severely damaged, with severe spatio-temporal disorientation, when long-term memory becomes impaired later [1][2]. When different hypothesis may explain the causes of AD, there is no known validated pharmacologic therapy. Nevertheless, the research effort present promising results. Non-drug therapies are adapted to flatten or compensate the AD effects by stimulating cognitive and sensorimotor activities such as music practice and dance, which convene neural plasticity processes [3]. Playing and listening to music all together stimulate social forms of cognition, emotion and participate to re-entrainments of implicit and procedural memories affected by AD [4][5][6].

In the context of longterm care centres (EHPAD), these recent results encourage group musical activities led by music therapists in a variety of non-drug therapies for AD. Nevertheless in such a longterm care context, musical activity may be too rare and the benefits for each victim of AD are difficult to evaluate, particularly when the latter may present different forms of dementia.

As a geriatrician may observe in his own practice in such a context, too little attention is generally given to all sonic interactions which include not only music but various oral communications, noises, audio productions... and silences, all that actually defines the standard soundscapes of a living and social place in a hospital environment.

Moreover, such sonic interactions are able to recall facts from subjects' trivial (long term) memory, and reach the subcortical circuits spared by the disease, in particular those that are related to emotion as well as hearing.

In 2016, an experimentation by Dr Jeannin with AD victims showed that some peculiar sounds can *help recall long-term memories and emotions, when the listener's performance shows no relation with standard cognitive tests such as MMSE and NPI*. Dr Jeannin also demonstrates how long such an experience of auditory memory with simple sounds implies a knowledge of the biography of the listener : this biographical knowledge may facilitate the activity of caring by, for instance, sparking some *conversation* [7].

Therefore, some hospital's inner sound environment may not be well adapted to AD victims : during the day, loud radio or TV-shows play for a long time in individual or collective rooms and can make a continuous flow of informations that has mostly become incomprehensible in the studied cases of dementia. At night, wanderings are not rare in deep silence with no landmarks, no surprise, nowhere to go when the way back may be already lost in anonymous and closed halls, etc.

When sonic interactions may focus on the caring activity in a clinical perspective, a question may be: how would sound directly take part in AD caring ?

It is with this prospect that the Grégoire Direz Residence (EHPAD) in Mailly-le-Château has welcomed, since July 2017, a sound design research action in their unit specially equipped for AD victims. The hospital, its geriatric physician (Dr Pierre Jeannin), and its care team work together with a gerontologist (Prof. France Mourey[1]) on scientific aspects while I work on conceiving and carrying out the sound device on location: the « Madeleines sonores ».

To be pragmatic, we presume that from a functional and systematic point of view, not only music but every sound may involve various cognitive and emotional circuits [9][10]. An hypothesis would be that as soon as emotions and long-term memory still strong in later stages of AD (alexithymia appears late), not only music but any sound, if well chosen sounds, may help the caring activity [7] for AD : being *in situ* (i.e. on location), attention, social processes as well as emotion – itself able to move us – can be called upon as much through the music as from a variety of familiar or merely recognized sounds. Where, when and how sounds occur seem to be relevant questions to AD and dementia in a longterm health care centre.

Finally, one may consider that a continuity exists between "music" and "non-music" or "noise", which vary according to the human society and its history. According to Luigi Russolo, Edgar Varèse, John Cage, Pierre Henri, Muray Schafer and others, a silent rest, a siren and other urban or natural soundscapes may be considered as "music" in certain conditions. Anthropologists and musicologists have shown a very large variety of music productions in relation to humans, animals, nature or surnatural beings.

---

[1] Gerontologist, at Cognition, Action and Sensorimotor Plasticity Lab., UMR Inserm Unit 1093 - Université de Bourgogne, Dijon (France).

## 2 Designing Soundscapes for AD victims

Pragmatically, we suggest designing virtual soundscapes adapted not only to different stages of dementia due to AD, i.e. to different mind states and beliefs, but also to some of the physiological effects of aging, such as a reduced visual field, hearing loss (or blocked ears), altered movement control, relative perceptions of time, etc. Starting from my own experience, from philosophical and ecological aspects of soudscapes from philosophical and theoretical frameworks devoted to soundscapes and mental health [8][9][10], and from rare references to some experiments on sound and soundscapes for AD victims [11][12], I decided at first to chart the time and space of this very particular place and, then, to focus on anxiety phenomena that a sound environment may generate in certain conditions, particularly with elderly people.

After my first observations done for nights and days in the Alzheimer's Unit at the EHPAD under the supervision of Dr Jeannin and of the caring team, I decided at first to design the common areas of the Unit, where people spend place most of their time standing, moving, doing some activity or resting, with the care team.

The first sounds I had to produce had to obviously be congruent with the physical environment: big rooms, corridors, impersonal furniture, windows overlooking beautiful gardens and, in the distance, a village ... They also had to be congruent with the *immediate* perception and beliefs the residents may have, residents who are affected by (various forms of) dementia but who also bravely bear the various marks of now ancient times! To answer these multiple deteriorations of perception (vision, audition, proprioception), I deemed necessary to start with reinforcing what already existed to increase perception, thus "reality" in some ergonomical perspective. This is the context in which, afterwards, sounds were supposed – tactfully – to encourage each one's imagination of former memories, collectively as well as individually, ordinary as well as extraordinary : not only with sound samples played out of their context, but sound effects.

A great variety of sound effects may partake in the staging of different sound realities and their modelling. With that prospect, I suggest understanding the notion of sound effect with an emphasis on a "logic of meaning", between the cause and the event itself, "in a dimension that is altogether that of event and of situation", through the "demonstration of a phenomenon that supports the [assumed] existence of the object" [13]. Therefore, the staging of this particular hospital space, that has become the canvas of a "film for the ear", must be as much part of the scripting as of the content of the generated sound flows. By the way, the digital audioprocessing used in the actual disposal are not numerous: some peculiar synthesisers and peculiar sound players with volume controls, fading, filters and a reverb. On the contrary, the logical and stochastic effects are much richer and are the subject of the original computer coding (see below Section 4).

In the map above (see Figure 1), sounds produced from realistic sources from outside are supposed to be welcome in such a hospital's closed space : at its extreme borders, synthesized sounds of (virtual) bells give night and day to the listeners the (real) hours... or a landmark, as well as a short but regular game for cognitive stimulation (counting the strokes). In the patio, a granular audio synthesizer produces a continuous but varying waterfall sound, congruent with the beautiful but silent view, when the windows cannot be opened for security reasons.

This very first signals appear to be strong enough as landmarks to allow some algorithmic variations, depending on a mix between the actual and virtual place, on both the weather and the rhythm of the caring activity : when the sound quality of the waterfall or of the bells can change according to the outside temperature, or to the amount of humidity in air, their loudness intensity varies according to the care activity as well as to the cosmological events. Periods of days and seasons are also marked by a variety of geophonic sounds such as rain and wind which vary according to the actual weather... On the biophonic side, animal populations and their audio activity also change according the actual season, weather and ethological data. In the same way, the biophonic production is in relation with the precise location of the broadcasting of sound: birds that might come – whether in reality or virtually – to the waterfall in the patio are not the same ones as those that can be heard in the little living room on the North side, neither are they the same as those that can be heard in the big living on the South side.

The time-length of the animal sound sequences is approximately twenty minutes long, depending on weather conditions. Their intensity and density, therefore the duration of silences, depend on season and weather. Thus different tones continuously blend while they « naturally » diversify, as the algorithm gives the sound of birds in the rain, that would be more often crows in winter and tits in spring. When the sonorous broadcasting through loudspeakers is allowed for by the layout, audio spatialization effects may be produced, with moderation, to render movements coming from the wind, trees, birds, mammals...

Sounds that are distinctive to human activity, more variable are arbitrary, are broadcast in other (more infrequent) places of the common space. For example, in the area of the dining room sounds may suggest meal preparation. In this case, aiming at realism, and especially at respecting a multimodal congruence (here audition vs. taste), the choice of sound samles can, eventually, be made consistent with the menu, and the sound of a pressure cooker made become more frequent in winter than in spring: this implies, therefore, collaborating with cooks !

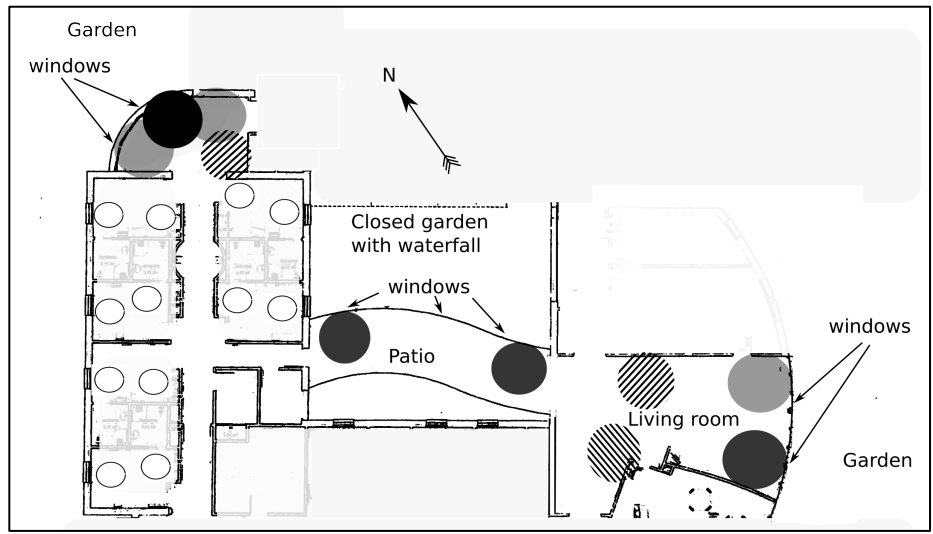

**Figure 1.** Layout of the loudspeakers in the common spaces of the Alzheimer's Unit: 1) the broadcasting points of the landmarks: bells and waterfalls (*black circles*), can also play geophonic and biophonic sequences ; 2) the broadcasting points of sounds for geophony or biophony and occasional human sonic activities from outside (*grey circles*) ; and 3) the broadcasting points of human audio activities, mostly from inside (*dashed circles*). The bedrooms have two broadcasting points each (*smaller empty circles*) : only six of them are shown on this map.

Extreme attention is given to the sound pressure intensity. I adapted the sound production with a rather unusual but conscious concern for the continuously exposed audience. Calibrated measures of acoustic pressure are underway.

In the common spaces, that are the noisiest, the acoustic pressure produced is unevenly distributed. Periods of silence (the acoustic pressure being of roughly 30 dBA) are frequent, except in the patio that overlooks the inner garden and its waterfall whose sound is continuous, though variable during long periods depending on time and weather conditions (minimum pressure of roughly 40 dBA). From the little living room on the north side, some sounds are broadcast loud enough to be heard as we arrive from the corridors that lead to it. In this more isolated place than the others, that leads to meditation or strolling, periods of silence alternate with active periods (with an acoustic pressure of 70 dBA and peaks at 80 dBA).

Until now, as is confirmed by the written notes taken by the care team, voiced complaints have never been about sound pressure, but rather about their eventual existence and pertinence. On the contrary, it even seems as if, with time, sounds tend to not be noticed consciously, at least by the health care workers. In agreement with the care team, the general sound pressure has even been reinforced with experimentation, in particular to compensate the eventual hearing deficiencies of the residents.

As recommended by music therapists [14], the general shape of a sound sequence, aside from the sound landmarks, for example the chirping of a bird, or horses passing, etc. follows as much as is possible a U-shape for about 20', most often followed by a period of silence.

Insofar as we offer to encourage mnesic recalling and, in order to achieve that, to resort to attention cognitive processes, we suggest also resorting to a form of ABA' shape in which the sound shape A would tend to ressemble a call. It can be noticed, as a matter of fact, that the U-shape, that can also change into an inverted "J", conjures up a kind of returning (if only that of an increasing tempo), like the ABA' shape.

Each broadcasting point being independent, several sequences can partly overlap in various places. The number of simultaneous voices in the same location remains, however, voluntarily limited so as not to saturate the sound space, without, for that matter, preventing more or less fortuitous sounds meeting. Silences also allow, depending on what sounds are conjured up by the context, various effects that can concern sometimes a feeling of expectation (calls), at other times relief (response), mirroring the shape of a rhythm.

When the soundscapes of the common spaces take their inspiration from an as common reality, the broadcasting of soundscapes in the private spaces (bedrooms) is much more personalized. The scripting strategies there are somewhat different since the idea is to conjure up the residents' old memories while avoiding to cause confusional states that would take place in particular moments that the care team would help define. The choice of sounds and their staging here has to answer the congruence with a reality perceived in a state of dementia – which needs to be diagnosed in the best possible way – as well as the life story of the listener. We also have to admit that such diagnoses are not often enough made, with regards the evolution of the disease. Different protocols, based on the above-mentioned principles, are being considered.

Some simple and repetitive sounds, that conjure up a bygone past, may have a soothing effect: an initial observation, along with the care team, showed that the warm, slow and regular sound of a grandfather clock's pendulum, slowed down to an every-two-second impulse, seems to ease the sleeping phases and allow for a better sleep. Such observations are still, at this point of experimentation, to be confirmed experimentally: the rooms are in the process of being empirically and progressively fitted with sound. We hope that, in the long run, such an experimentation may enable to elaborate a precise questionnaire, based – for example – on a guided listening of sounds which are defined beforehand, that better allows to characterize each resident's sensitivity to different sounds.

This personnalized aspect of conception, the most delicate, is in the course of being experimented and will progressively be deployed.

## 3 Soundscapes for caregivers

Obviously, some sounds may be difficult to play with. In some conditions, with dementia, when the perception of some sounds imply, as a reflex, the viewing of their origin, their separation from their real or imaginary cause may create confusional states: to avoid such a case, the help of a caregiver is useful to explain the origin of such a sound, to calm the person down and to finally have — as expected — some *conversation*.

Conceiving a sound environment and its dynamics, necessariy involves the health care team. Sound must not interfere with care, it must help and support it. We can therefore consider that sound may help the patient in directly adressing him or her while addressing the care team as well. With their assent and their complicity, oral conversations may be engaged in and sounds adopted.

For example, in the very first days following the beginning of the setup (in July 2018), nursing auxiliaries asked me for the new bells in the commons spaces never to leave: the opportunity and ergonomics of such a multi-functional sequence of a simple sound, repeated twice an hour (at 2' interval), 24 hours a day allows to mark a territory, to recall an hour common to all, to be the opportunity for a social exchange, and to become a training of different short-term memories, indeed to become part of spontaneous cognitive assessment, almost at any time. And slight periodic variations of the sound synthesis of bells can be sought for, for who might want it if only per chance. More or less shared knowledge, whether popular, personal or specialized, may be called upon on the subject of some animal sound production. Producing some insect sounds, or mammal ones, for example, has been the subject of discussions with the care team as regards their realism, the cridibility, or the emotional states they are likely to cause.

Soundification of virtual human activities may also take part in the organisation of care: for example, the sound accompanying meals may be realised in such a way that it more or less consciously prepares the residents to eat. This anticipation may be all the more appreciated as the residents themselves are incapable, because of their dementia, of knowing that it is *in fact* really time to eat... Such an effect is most certainly welcome when it contributes to the support of demented people in the absence of any other clue that would be obvious and familiar to them, in a hospital context.

The developpement of a closer feedback and control interface from the care team is contemplated. Conceiving such an interface would should directly integrate existing computer interfaces already devoted to caring. With that as a goal, research in psychosociology on the links between sonic environment and care has already began under the supervision of Prof. France Mourey. A first study of the effects of "Madeleines sonores" (the virtual soundscapes, the plan of action) on the relationship between caregivers and care receivers is based on collecting information during interviews with the various professionals linked with this project. The thematic analysis [15] of the interviews that were carried out have allowed to establish one (or more) questionnaire(s) that may characterize the representations and attitudes that the care team takes up faced with the sound plan of action[2]. We are considering that the use of such a questionnaire may allow to know the attitudes that come into play while applying the sound plan of action, this being to clarify the contribution in relation with the resident. Moreover, knowing how the care team sees this sound plan of action also partakes in evaluating its effects, in terms of contribution or benefits, with the residents.

---

[2]  I would like to thank Arnaud Bidotti, Master in Occupational and Organizational Psychology who led this study in March and April 2019 under the scientific supervision of Prof. Edith Salès-Wuillement, Laboratoire Psy-DREPI EA-7458 (Psychologie: Dynamiques Relationnelles et Processus Identitaires), Université de Bourgogne.

## 4 A Digital Architecture for Virtual Soundscapes

The soundscapes are systematically generated and mixed in real-time using a distributed architecture with a principal generator (see the « Soundscape Generator », in Figure 2 below) that can be described as a server hosting the sound database and a set of routines periodically executed by its operating system scheduler[3]. These routines, written in Lisp language and Unix-like shell scripts, compute different kinds of mapping, empirically developed, relevant to the environmental data as modelised, to its inner virtual population and to individual data (biographies). They finally control 24 on-board computers[4] fixed in the ceiling of the common and private spaces and hosting their own digital signal processing (DSP) and loudspeakers (the « Players », in Figure 2).

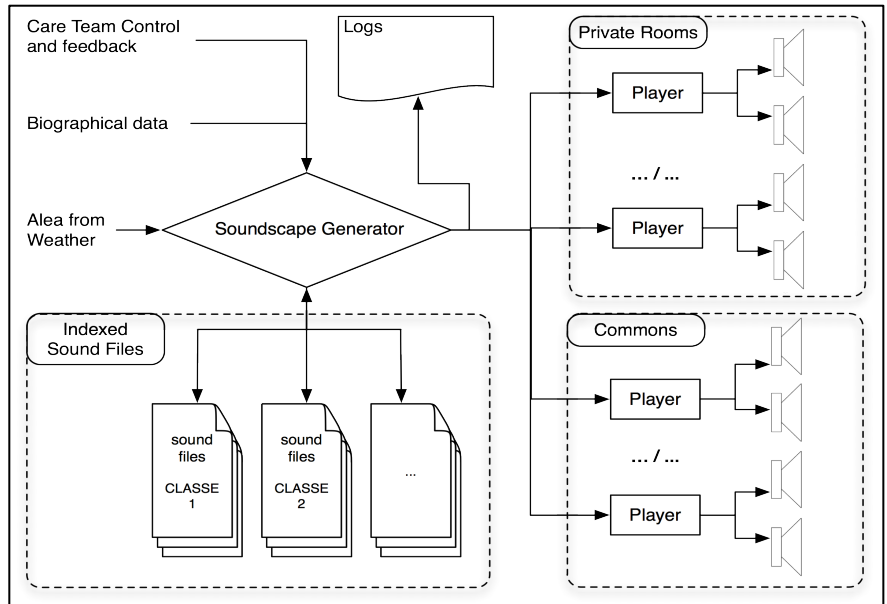

**Figure 2.** General principle of the *Madeleines Sonores*

All the soundscapes are generated as a remix of short sound samples (lasting from a few seconds to a few dozens of seconds), or using sound synthesizers models adapted from my own experience [16] and from [17]. The sound samples, specially recorded or adapted from open sound collections to represent various places, activities, animals and human societies from the 40's, 50's, are retrieved from a large database with thousands of items indexed by myself. Periods of silence between sound samples or synthesizers are more or less short and always irregular. They define the density of virtual events that occur depending on each resident's sensitivity and the general context: according to daytime periods, the year, the care provided, etc.

---

[3] Cf. cron programs in Unix-like operating systems (see: https://en.wikipedia.org/wiki/Cron ).

[4] All computers operating systems are GNU-Linux based. The on-board ARM based computers are Raspberry Pi3-B+ with audio boards IQaudio Pi-DigiAMP+ and in-ceiling 6.5" loudspeakers QI 65C/St by QAcoustics (70 Hz to 16 kHz ± 3 dB, up to 90 dBA max pressure level). The DSP is built using Pure Data open-source software [18]. The communications are based on TCP/IP protocols: secure-shell (SSH) for asynchronous sound transfers, UDP for real-time controls.

In the individual rooms, the soundscapes are rendered using two equivalent monophonic loudspeakers placed across the room, one close to the window and the other facing the bed. The activation of the soundscapes entirely depends on the explicit approval of whom concerned: the individual, himself, herself or their family when he or she cannot agree for clinical reasons and, obviously, by the caring team (doctors, psychologists, nurses).

All hi-level instructions, play-lists, sound filenames, pressure levels, etc are systematically locally logged. The overall disposal can be remotely controled and monitored thru a standard secured connection (SSH) on internet, using any Unix-like terminal. Confidentiality is ensured in particular by the entire anonymity of the digitized data.

## 5 Discussion

The work presented here is only the beginning of a long conceiving process in an emerging activity within the field of sound design, mental health and care. Its actual state is an action-research led in the fields of sound art, human and health science, and technology.

Our first observations are encouraging: the behaviour of part of the residents suggest that not only the residents appreciate the variety of the new sound environment but they tend to place themselves close to the audio sources to recollect or to meditate, whether consciously or not (see Fig. 3 below). The questionnaire we developed should make it possible to make care givers aware of the effects of sound environment on AD victims.

After such a first research that was necessarily empirical, considering the innovative characteristic of the project, applications respecting a truly experimental protocol have to be considered.

Optimizing the algorithms that was developed for that purpose still has to be done, as the deployment of the plan of action is being ratified. Particular attention should be focused onto indexing the sound samples insofar as they should enable, not only to identify the origin, the nature and the quality of sounds, but also, beyond their psycho-accoustic properties, the emotions and reactions they are liable to cause, as well as the broadcasting methods (context). In fact, we are contemplating the fact that, ultimately, the care team may report the reactions they provoke thanks to an interface that remains to be created. Such an interface would then enable to directly partake in the experimental phase of the plan.

The whole plan of action is for the moment operating as a multimedia work of art (sound and computer) that falls within the altogether artistic, scientific and clinical perspective that federates different social and economic players in a humanistic dynamic of research and innovation.

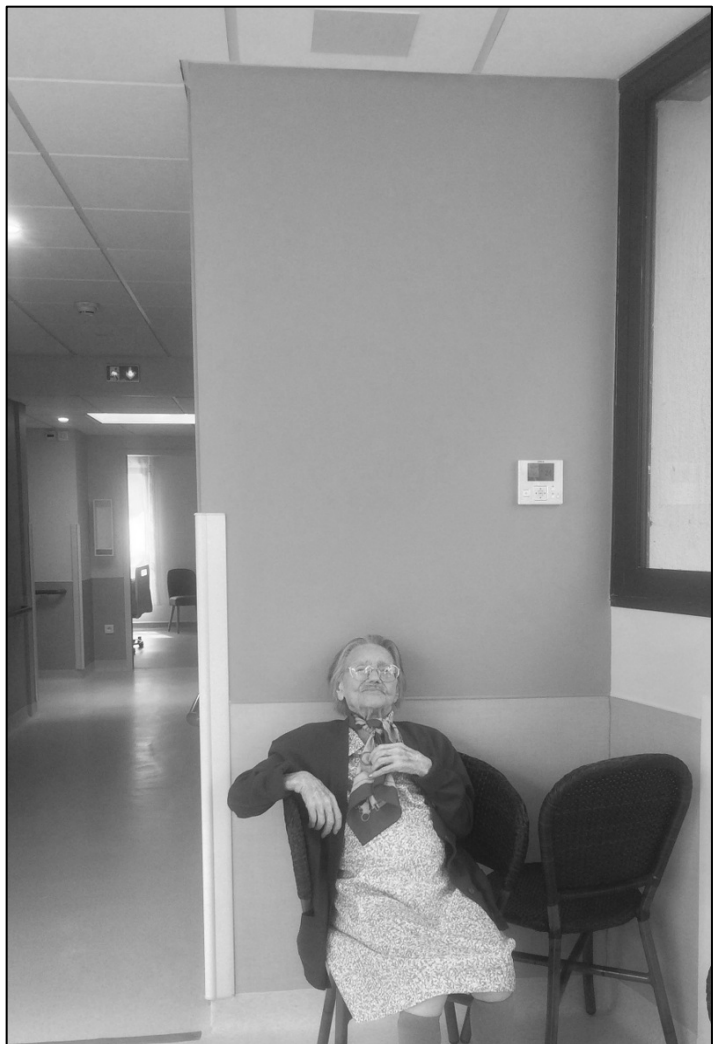

**Figure 3.** A resident is meditating at the closest place to the audio source mounted at the ceiling.

## Thanks

I am particularly thankful to Dr Pierre Jeannin for his benevolence, his logistic and voluntary support, and of course to the charity Castel-Mailletaise that welcomed this project, and the care team that is closely involved in it. My thanks also go towards France Mourey for her support. And I am grateful to Claire Webster for emending and translating parts of this article.